\documentclass[aps,pra,reprint,groupedaddress,twocolumn]{revtex4-2}
\usepackage{graphicx}
\usepackage{xcolor}
\usepackage{amsmath}
\usepackage{mathtools}
\usepackage{amssymb}

\begin{document}

\author{Giuseppe Ortolano$^{1,2}$}
\author{Ivano Ruo-Berchera$^3$}
\author{Leonardo Banchi$^{1,2}$}
\affiliation{$^1$ Dipartimento di Fisica e Astronomia, Università di Firenze, Via G. Sansone 1, I-50019 Sesto Fiorentino (FI), Italy}
\affiliation{$^2$ Istituto Nazionale di Fisica Nucleare, Sezione di Firenze, via G. Sansone 1, I-50019 Sesto Fiorentino (FI), Italy}
\affiliation{$^3$Quantum metrology and nano technologies division, INRiM, Strada delle Cacce 91, 10153 Torino, Italy}

\title{Chernoff Information Bottleneck for Covert Quantum Target Sensing}

\begin{abstract}

  The paradigm of quantum metrology and sensing aims to identify a quantum advantage in precision at a fixed energy of the probe state. However, in practice, employing high-energy classical probes is often simpler than leveraging the quantum regime. This is not the case of covert sensing scenarios, where detection must be performed while avoiding to be discovered by an adversary, because increasing energy unduly facilitates the adversary. In this paper, we introduce a general framework to assess the quantum advantage in covert situations based on extending the information bottleneck principle to decision problems via the Chernoff information. We demonstrate how entangled photonic probes paired with photon counting significantly outperform classical coherent transmitters in covert detection and ranging, often representing the only option for secrecy. Thus, our work highlights the great potential of integrating quantum sensing into LiDAR and Radar systems to enhance covert performance.
\end{abstract}
\maketitle

\section{Introduction}\label{sec:intro}
Quantum sensing \cite{Giovannetti_2004, Giovannetti_2011,Degen_2017} is a highly active field of research that in recent years has offered many promising protocols with great potential for technological applications \cite{Genovese_2016, Pirandola_2018, Berchera_2019}. Among those are quantum ranging \cite{Zhuang_2021,Zhuang_2022b, Ortolano_2024} and detection \cite{Lopaeva_2013, Shapiro_2020, Lee_2023,Karsa_2024, Torrome_2024}, which have attracted a great amount of interest and have been investigated in great detail since the original proposal of quantum illumination \cite{Lloyd_2008, Tan_2008}. This is due to the potentially groundbreaking advances that quantum resources can offer to LiDaR and radar systems \cite{McManamon_2012}, and the consequent widespread applications. 
 
In the path to practical implementations some criticalities remain open. Quantum information inspired protocols exploit a sequence of two modes, named signal and idler, in an entangled state, where the signal's modes are addressed to the target region while the idler's modes are retained locally for a final joint measurement with the back-reflected signal.
One restriction is that the quantum advantage is found in a range of parameters which is seldom the one used in practical scenarios, namely the very low energy of the single mode irradiated towards the target and, at the same time a high thermal background mixing with the signal. As a consequence, a large time-bandwidth product, i.e.~large number of modes of the quantum source, is needed to achieve a meaningful signal-to-noise ratio, while in the classical case this is not a requirement \cite{shapiro2020quantum,sorelli2021detecting}. 

A more natural application of quantum schemes can be found in covert sensing \cite{Tham_2024,Bash_2017,Gagatsos_2019,Tahmasbi_2021}, which explores the situation where probing has to be performed while avoiding detection by an adversarial party. The requirement of covertness prevents the probing party from arbitrarily increasing the energy  to reach a better performance, meaning that an energy constraint is inherent to the problem.

Aside from the aforementioned constraints, quantum target sensing also requires a technologically demanding quantum memory to store the idler modes and phase-sensitive joint measurements. Some of those requirements can be strongly relaxed if the quantum advantage is sought at a fixed type of measurement. In the optical regime, phase sensitive measurements are not efficient in realistic contexts of remote sensing, affected by speckle noise from rough target, diffusing media and turbulence \cite{mcmanamon2012review}. In fact, time-of-flight evaluations through intensity/photon counting measurements are nowadays the state of the art for Lidar applications, also prompted by the development of picoseconds time-resolved single-photon detectors \cite{morimoto2021megapixel,tachella2019real}. In this case, one could simply avoid to involve quantum memories or unpractical interferometric setups.

Motivated by the above challenges, 
here we develop a general approach to covert sensing based on the concept of Chernoff information bottleneck. 
The information bottleneck was  historically introduced via entropic quantities
and has a long history of applications \cite{Tishby_1999}. In its standard formulation with 
two correlated random variables $X,Y$, the information bottleneck method aims to find the best 
compression of $X$ that still allows for an accurate reconstruction of $Y$, thus quantitatively  introducing 
a trade-off between compression and accuracy. 
Quantum generalizations of the information bottleneck principle have been applied to diverse 
problems, ranging from quantum communication \cite{datta2019convexity} to quantum machine 
learning \cite{banchi2021generalization,hayashi2023efficient}. 
Here we consider a different formulation based on the Chernoff information that is more suited to decision problems,
with the goal of quantitatively optimizing the trade-off between covertness and sensing ability. 
In physical terms, what is ``compressed'' here is the energy of the probe. 

As a relevant application, we evaluate the quantum advantage for cover target ranging with photon counting measurements,  which are particularly suited to describe this technology in the optical regime. 


\section{Covert Target Sensing }\label{sec:covert}

\begin{figure}[t]	
  \includegraphics[width=0.5\textwidth]{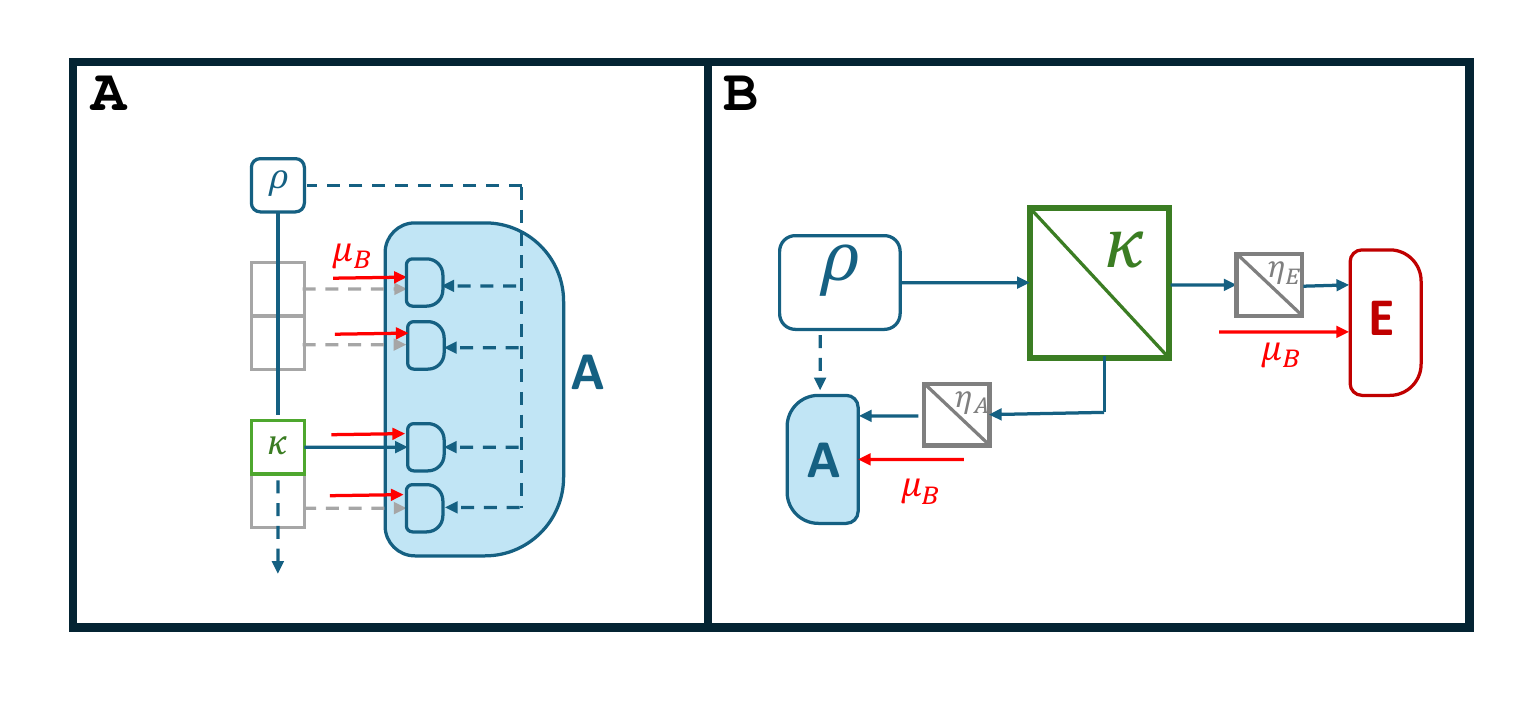}
	\caption{\textit{Ranging and detention schemes}. \textbf{A}. Ranging is performed by Alice that sends a probe returning in one of $m$ possible slots denoting different positions of the target. A decision is made after measuring the returning signal jointly with the unavoidable background noise. \textbf{B}. The covert sensing is performed by assuming that all (up to the collection efficiency) the signal that does not return is collected by an adversary performing passive measurement to detect Alice. }\label{fig:Mod}
\end{figure}

We focus on target sensing tasks by one party, 
labeled as Alice (A), under covertness conditions \cite{Pace_2009,Bullock_2020} against an
adversary, denoted as Eve (E). 
Example tasks include target detection \cite{Tan_2008, Shapiro_2020, Karsa_2024}, which can be modeled as a binary decision problem to assess 
whether a target is there or not, or target ranging \cite{Zhuang_2021, Zhuang_2022b, Ortolano_2024}, 
where the task is to assess the position of the target among a discrete set of choices -- 
see Fig.~\ref{fig:Mod}\textbf{A} for reference. Both tasks find applications in the design of quantum radar/LiDAR systems
\cite{Torrome_2024,Maccone_2020}.
Alice performs the sensing  task  using a probe, sent
by a transmitter in a given state $\rho$. The probe interacts with the target, which 
is assumed to be a, typically faint, reflective object, so it is modeled as an
optical beam splitter of reflectance $\kappa$. After the interaction with the
target, the probe state comes back to a receiver, where a measurement is
performed. According to the nature of the problem, we consider an unavoidable
multi-thermal background. 

For the adversary, on the other hand, the  problem is always the binary decision on the Alice’s presence.
This is performed by Eve by collecting all the signal that is not reflected
back to Alice, i.e.~the other output of the beam splitter modeling the target,
see Fig.~\ref{fig:Mod}\textbf{B}. Both parties receive their respective signals after they
go trough pure loss channels, with transmittance $\eta_{A/E}$. These channels
define the collection efficiency and models environmental losses, as well as
the detection efficiency, that are not necessarily the same for Alice and Eve. We consider, for the sake of simplicity, Alice and Eve to operate in the same environmental background, e.g. in the same daylight conditions. One can show that the results are not qualitatively affected if the unbalance is not too great, under the condition $\mu_B\gg\mu$.

\subsection{Chernoff Information Bottleneck}\label{sec:ib}
 
Covertness can be defined in different ways \cite{Bash_2017,Gagatsos_2019}, and
the most common in the literature is to employ information theoretic quantities.
Specifically, some works have considered the relative entropy of the probe at
the adversary detection point \cite{Tahmasbi_2021}, with respect to a vacuum
state transmitter, that ideally in active sensing should gather no information.
The best performance in the covert tasks is then defined as the optimal
probability of error that can be obtained under the constraint of having this
relative entropy under a fixed threshold. The relative entropy is related to distinguishability \cite{Audenaert_2014} and sets bounds on the probability of detection  \cite{Ogawa_2000, Bullock_2020}.
Covertness in terms of a fixed probability of detection was formalized in Ref.~\cite{Tham_2024} and previously discussed in \cite{Bash_2017,Gagatsos_2019,Tahmasbi_2021}.

Here we adapt the latter approach, in particular we 
measure the performance of Alice and Eve's decision processes via the Chernoff information $\xi$ 
\cite{Nielsen_2013, Audenaert_2007}, which gives 
the asymptotic decay rate of the symmetric probability of error $p_{\mathrm{err}} \propto e^{-M\xi}$ in the limit of large repetitions (or large number of modes) $M$. 
We define two Chernoff informations, one for Alice's target sensing task, $\xi^{(A)}$, 
and another one to measure Eve's ability to spot Alice, $\xi^{(E)}$. 
More details are provided in the End Matter. 

We then define the \emph{covert information}, $I_{\text{C}}$, as the quantity:
\begin{equation}
I_{\text{C}}(d,\mathcal S):=\max_{\xi^{(E)}\leq d}\xi^{(A)},
\label{eq:IC0}
\end{equation}
where the maximization is over the probe states $\rho_0\in\mathcal S$, under the constraint that Eve's Chernoff information 
is limited by $d$. 
In other words, by formalizing the covert target detection problem as the maximization in Eq.(\ref{eq:IC0}), 
we implicitly define a \emph{bottleneck} on the related Chernoff information of
Alice and Eve. 
Different values of $I_{\text{C}}(d,\mathcal S)$ are possible depending on the chosen set of probe states. 

The problem \eqref{eq:IC0} defines a constrained optimization and its solution by KKT conditions
\cite{Kuhn_1951} can by found from the saddle points of the Lagrangian
$\mathcal{L}_{\text{C}}= \xi^{(A)}- \beta \xi^{(E)}$, with $\beta\geq 0$. 
In the Lagrangian formulation, the optimization is still with respect to the probe states 
in a given subset $\mathcal S$, 
but the dependence on the constraint $d$ is lost. 
Nonetheless, from the stationary points of the Lagrangian one obtains 
the optimal $\beta$-dependent rates $\xi^{(A/E)}(\beta)$ and can 
plot the parametric curve $(\xi^{(E)}(\beta),\xi^{(A)}(\beta))$. 
Such curve is monotonically increasing as a higher $\xi^{(E)}$ allows a higher 
rate $\xi^{(A)}$. Therefore, there exists a single value of $\beta$, that we call 
$\beta_d$,  such that $\xi^{(E)}(\beta_d)=d$. 
The optimal rate achievable by Alice under the constraint \eqref{eq:IC0} is 
then $I_{\text{C}}= \xi^{(A)}(\beta_d)$. The  covert information thus defines 
a curve $\xi^{(A)}=I_C(\xi^{(E)},\mathcal S)$ that links  the optimal rate
$\xi^{(A)}$ to the optimal, namely the largest, $\xi^{(E)}$.

Our choice of $I_C$ in Eq.~(\ref{eq:IC0}) is motivated  by the operational
meaning of the Chernoff information as the asymptotic exponential decay rate of the
probability of error, $p_{\rm err}\sim \exp(-\xi M)$ for $M\gg1$.  
Covert sensing can then be defined by two requirements 
\begin{align}
  \epsilon &:= e^{-\xi^{(A)}M},  & 1-\delta &:=e^{-\xi^{(E)}M}, 
  \label{eq:deltaeps}
\end{align}
namely that the probability of error for Alice is a small quantity $\sim \epsilon$, 
while the probability of being detected by Eve is $\sim\delta$ above random guessing, with $\delta$ another small quantity.
The covert information curve $\xi^{(A)}=I_C(d{=}\xi^{(E)},\mathcal S)$ then
links the errors in Eq.~\eqref{eq:deltaeps}.
Indeed, via Eq.~\eqref{eq:deltaeps} we get 
\begin{equation}
  \epsilon = e^{-M I_C\left(\frac{-\ln(1-\delta)}{M},\mathcal S\right) } 
  \approx e^{-\lambda  \delta^\gamma M^{1-\gamma}},
  \label{eq:scaling}
\end{equation}
where $I_C(d,\mathcal S) \approx \lambda d^\gamma$ is the expansion of the covert information
curve when $d\to 0$,
and $d\simeq \delta/M$.
If the covert information curve 
is approximately linear ($\gamma =1$), then covert sensing is possible only if $\lambda \delta\gg 1$,
irrespective of $M$, namely when the coefficient $\lambda$ is much larger than $\delta$.
When $\gamma>1$ cover sensing is impossible and increasing $M$ reduces the performance. 
On the other hand, if the covert information curve is sublinear ($\gamma <1$), then 
$\epsilon\to0$ whenever $M\to\infty$, thus ensuring covertness for any $\delta$. 

From the physical points of view, covert sensing requires the optimization of two quantities, 
the probe state and the number of modes $M$, given the two constraints in Eq.~\eqref{eq:deltaeps}. 
Operationally, when $\gamma<1$ it is always possible 
to find the minimum $M$ that ensures covert sensing by
inverting Eq.~\eqref{eq:scaling}, namely as 
$M\geq \left(\frac{\ln(1/\epsilon )}{\lambda \delta ^\gamma}\right)^{\frac{1}{1-\gamma}}$.
Once the $M$ has been chosen, it is then possible to find the 
parameters of the optimal probe, e.g.~the average photon number, by one 
of the equations in \eqref{eq:deltaeps}, introducing there the explicit expression of the corresponding Chernoff information. 
This argument outlines how the available number of modes $M$ and the average photon number
are  both important resources in covert sensing tasks.

Having clarified the role of the information bottleneck principle in 
ensuring covert sensing, we now analyse the calculation of 
the covert information $I_C(d,\mathcal S)$ in  \eqref{eq:IC0}.
It is often complex to explicitly compute $I_C(d,\mathcal S)$
for the most general class of probe states, namely when  $\mathcal S$ is the full Hilbert space.  
Therefore, we may fix the class of probe states that Alice can send, e.g.~squeezed states 
or coherent states, and compute the resulting Chernoff informations $\xi^{(A)}$ and $\xi^{(E)}$.
We will mostly focus on two possible choices, the ``classical'' set $\mathcal S_{\mathrm{C}}$ 
of coherent states and the ``quantum'' set $\mathcal S_{\mathrm{Q}}$ of entangled signal and idler states. 
We will claim quantum advantage whenever $I_{\mathrm{C}}(d,\mathcal
S_{\mathrm{Q}})>I_{\mathrm{C}}(d,\mathcal S_{\mathrm{C}})$ for the same $d$, namely 
whenever $\xi^{(A)}$ is strictly larger with entangled inputs than with coherent states, despite 
not exceeding the constraint $\xi^{(E)}\leq d$. 
Thanks to Eq.~\eqref{eq:scaling}, such advantage is asymptotically useful when the expansion coefficients of 
$I_C(d,\mathcal S_{\mathrm{X}})\approx \lambda_{\mathrm{X}} d^{\gamma_{\mathrm{X}}}$
satisfy $\gamma_{\mathrm{Q}}<\gamma_{\mathrm{C}}$, where $\mathrm{X}\in\{\mathrm{C},\mathrm{Q}\}$.
Stronger results are also possible: e.g.~if $\gamma_Q<1$, but $\gamma_C\approx 1$ and $\lambda_C=\mathcal O(1)$,
then efficient covert sensing is possible using quantum probes but impossible using classical strategies.
We foretaste that this stronger notion of advantage is possible in optical detection tasks.

Having clarified the formalism and the figures of merit, we can now focus on the target ranging protocol, and compare the performance of quantum and classical probe states.

\section{Application: Covert Target Ranging}\label{appl:target ranging}

The problem of target ranging can be expressed as a $m-$hypoteses test, as in Eq.~\eqref{eq:alice hyp} 
with $
\rho_{j}:=\rho_{\kappa}^{(j)}\otimes \bigotimes_{i\neq j}^{m-1}  \rho_B^{(i)},$
where the tensor product structure denotes the $m$ slots of the state, 
$\rho_B$ is the state of the background and
$\rho_{\kappa}=\mathcal{E}_{\kappa,\mu_B}(\rho_0) $ is the returning signal
after the interaction with the channel and the mixing with the background,
described by the quantum channel $\mathcal{E}_{\kappa,\mu_B}$. Our analysis
focuses on the optical regime and we choose to model the interaction as
proposed in Ref.~\cite{Ortolano_2024}, i.e.~a pure loss channel of reflectance $\kappa$ followed
by an incoherent mixing with a large number of background modes having a total number of
photons $\mu_B$. Here, the idea is to describe a situation in which the modal structure of the returning signal is unpredictable and non-stationary, due to target and environment scattering, and the detector integrates over many spatial-temporal modes in order to maximize the collection efficiency. In this varying multimodal scenario it is also natural to expect phase information to be hardly preserved.  So, here we consider photon counting measurement, motivated by the fact that interferometric approaches would be not useful.

\begin{figure}[]	\includegraphics[width=0.5\textwidth]{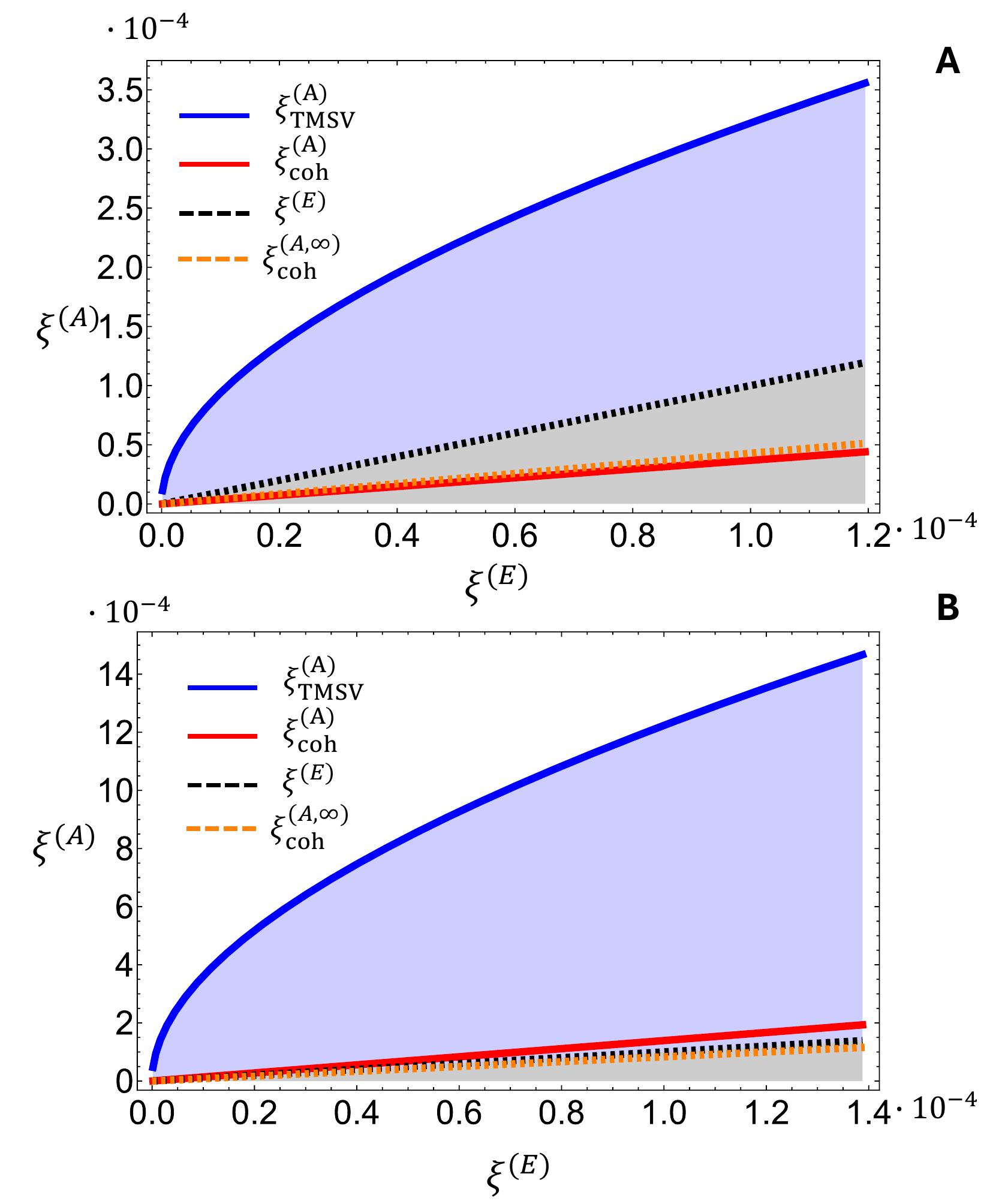}
	\caption{\textit{Sensing-covertness trade-off.}  We plot the pair $\{\xi^{(E)},\xi^{(A)}\}$ for the TMSV (blue) and coherent (red) probe. The dashed orange line represents $\xi_{coh}^{(A,\infty)}$, characterizing the alternative classical strategy described in the End matter. The gray area denotes the region in which effective covertness (see main text) is allowed. The blue and red regions denote quantum and classical achievable sensing respectively. For panel \textbf{A} we set $\kappa =0.3, \mu_B=10, \eta_A=\eta_E=1$. For panel \textbf{B} $\kappa =0.2, \mu_B=1, \eta_A=1$ and $\eta_E=1$.  }\label{fig:2}
\end{figure}

\begin{figure}[]	\includegraphics[width=0.5\textwidth]{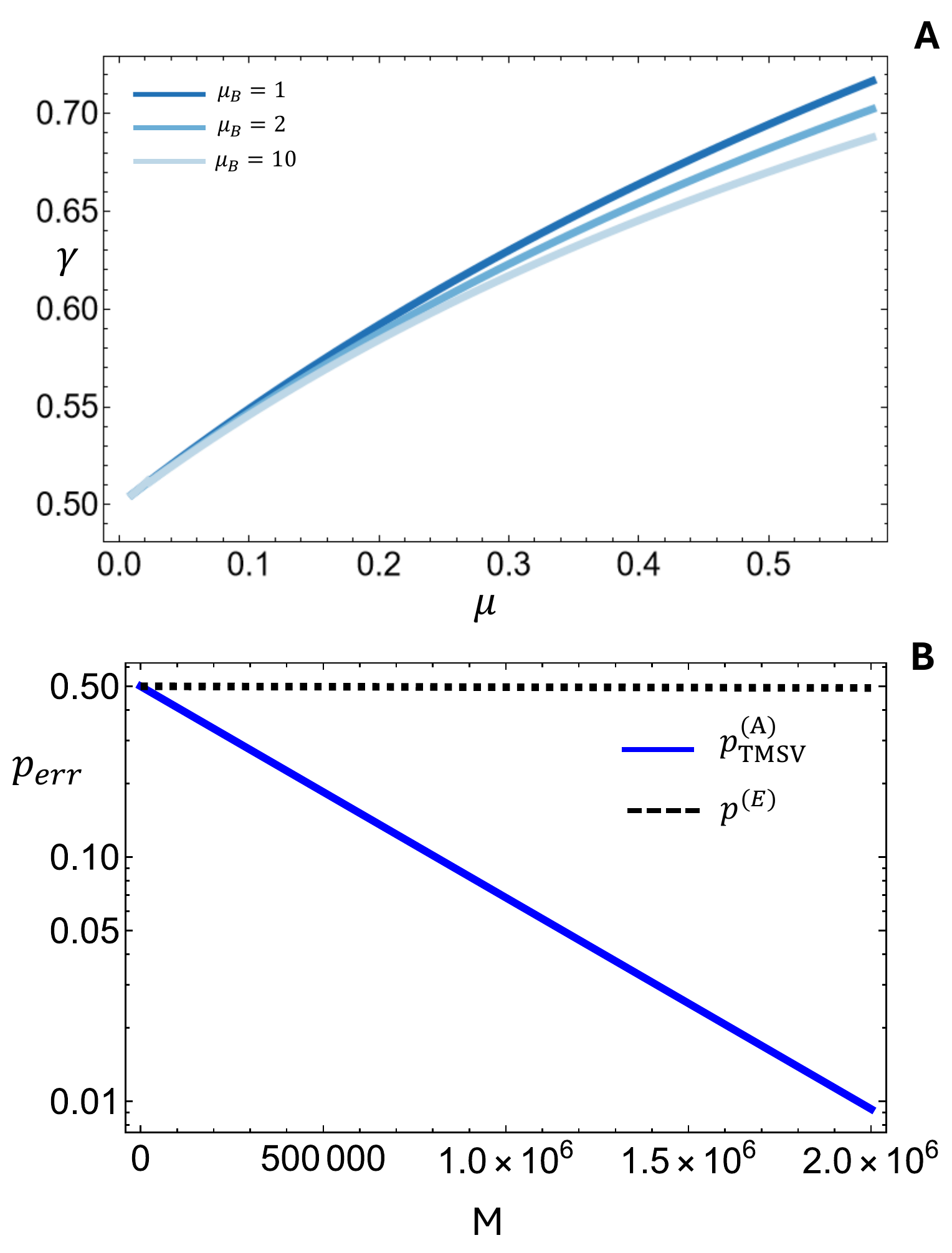}
	\caption{\textbf{A}. \textit{Chernoff information scaling.} We plot the parameter $\gamma$, defined in the main text, for the quantum probe as a function of the mean number of signal photons $\mu$ at different values of background. The parameters are $\kappa =0.2, \eta_A=1$ and $\eta_E=1$. \textbf{B}. \textit{Asymptotic probabilities of error.} Asymptotic probabilities of error (in log scale) in the limit of a large number of modes $M$. The parameters are the same of panel \textbf{A} and we fix $\mu_B=10$ and $\mu=0.001$.}\label{fig:3}
\end{figure}

\subsection{Sensing-Covertness Trade-off}
We investigate the quantities $\xi^{(A)}$ and $\xi^{(B)}$, in different parameter settings -- more results 
can be found in the End Matter. 
We compare in covert target ranging the performance of a specific ``quantum'' state of light with the class of  ``classical'' probes \cite{Mandel_1995}, i.e. convex combinations of coherent states. In practice, the optimal classical benchmark is achieved by a probe in a coherent state $\rho_{\rm cla}=|\alpha\rangle\langle \alpha|$ with $\mu=|\alpha|^2$ mean photons, having a Poisson photon number distribution.

At optical frequencies, the background is strongly multi-mode, each mode occupied with a low mean photon number, thus the photon-counting statistics can be approximated by a Poisson distribution \cite{Ortolano_2024}, with mean $\mu_B$. Alice's performance in the ranging task is, according to Eq.(\ref{eq:CITR}) \cite{Ortolano_2024}:
\begin{equation}
\xi_{\rm cla}^{(A)}=\kappa\eta_A\mu+ 2\mu_B-2\sqrt{\mu_B} \sqrt{\mu_B+\kappa\eta_A\mu}. \label{eq:CIA}
\end{equation}
On the other hand, Eve' performance is always given by:
\begin{equation}
\xi^{(E)}\approx \frac{1}{2}(1-\kappa)\eta_E\mu+ \mu_B-\sqrt{\mu_B} \sqrt{\mu_B+(1-\kappa)\eta_E\mu}, \label{eq:CIE}
\end{equation}
where the approximation holds true for a high background w.r.t.~the signal photons, $\mu_B\gg\mu$. This approximation steams from the fact that in the regime $\mu_B\gg\mu$, the optimization defining $\xi$ (see End Matter), is solved for $\alpha\approx 1/2$ \cite{Nielsen_2013, Tham_2024, Ortolano_2024}. Furthermore we can expand Eq.(\ref{eq:CIE}) for $\mu/\mu_B\ll 1$ to show the scaling $\xi^{(E)}\approx \frac{1}{4}\eta_E^{2}(1-\kappa)^{2}\mu^2/\mu_B$, expected for covert analysis \cite{Bask_2012,Bloch_2016}.
Note that, in this regime, $\xi_{\rm cla}^{(A)}$ has the same scaling in $\mu$, so that $\xi^{(A)}=\mathcal O(\mu^2)$ and $\xi^{(E)}=\mathcal O(\mu^2)$. This means that $I_C$ is approximately linear, $\gamma\approx1$ in Eq.~\eqref{eq:scaling}, thus efficient covert sensing is not possible. 

Having defined the classical benchmark, we compare it with a quantum probe $\rho_{\rm qua}=\left(|\mathrm{TMSV}\rangle \langle \mathrm{TMSV}|\right)^{\otimes R}$ composed of a collection of two-mode squeezed vacuum states, where $|\mathrm{TMSV}\rangle=\sum_n \sqrt{c_n} |n\rangle |n\rangle $ and $c_n$ is the  probability function of a thermal distribution with mean $\mu_0$. Using a collection of $R$ copies rather then a single TMSV states means that the marginal photon number distribution will be a multi-thermal one with $\mu=\mu_0 R$ mean photons. In the limit of large $R$ and small $\mu_0$ at fixed $\mu$ the distribution will be Poissonian. We choose the quantum transmitter in this regime as it gives an advantage in a wider region of parameters in ranging \cite{Ortolano_2024}, as well as mimicking better the background distribution in our model. Note that this also means that amplitude modulation is not necessary to enhance covertness. The Chernoff information of the quantum transmitter can be analyzed numerically and we denote it as $\xi_{TMSV}^{(A)}$.

Note that $\xi^{(A)/(E)}$ are monotonically increasing functions of $\mu$. Consequently, once $d$ is fixed, Eq.(1) is solved on $\xi^{(E)}=d$, that in turn determines $\xi^{(A)}$. As detailed in the previous section, the function $I_C$ still carries non-trivial information, as we show in Fig.~\ref{fig:2} where we report a parametric plot of the pair $\{\xi^{(E)},\xi^{(A)}\}$, often referred to as relevance-complexity in the information bottleneck formalism.  The blue line represents Alice's Chernoff information with a quantum probe as a parametric function of Eve's Chernoff information, that increases with the number of probe photons, while the red one refers to a classical probe. The black dotted line serves as a reference denoting the condition $\xi^{(E)}=\xi^{(A)}$. For TMSV probes (blue line), the covert information curve $\xi^{(A)}(\xi^{(E)})$ always behaves sublinearly and,
due to Eq.~\eqref{eq:scaling}, efficient covert sensing is achievable.
On the other hand, for coherent state sources (red line) the covert information is linear with an $\mathcal O(1)$ slope, corroborating the above analysis that covert sensing is not possible, also obvious from the fact that in this parameters setting $\xi^{(A)}_{cla}<\xi^{(E)}$.  Fig.~\ref{fig:2}.\textbf{B} shows the same plot with a reduced efficiency for Eve, so that $\xi^{(A)}_{cla}>\xi^{(E)}$. Even in this case, however, the linear scaling of the covert information prevents efficient covert sensing (see also Fig.\ref{fig:4}.\textbf{B}).
The orange dashed line refers to the figure of merit $\xi_{coh}^{(A,\infty)}$, characterizing 
the performance of a single coherent probe containing all the energy, $M\mu$, as detailed in the End Matter.
In Fig.(\ref{fig:3}.\textbf{A}) we plot, for the quantum probe, the parameter $\gamma$, introduced in Eq.(\ref{eq:scaling}), and computed as $\gamma =\text{d} \log{\xi^{(A)}_{TMSV}}/\text{d} \log{\xi^{(E)}}$, where we use the same expansion as in Eq.~\eqref{eq:scaling}. The plot also compares different backgrounds, showing how the magnitude of $\mu_B$ has relatively minor relevance, as long as the condition $\mu/\mu_B\ll1$ is met. As $\mu\to 0$ the parameter $\gamma$ approaches $\gamma\approx 1/2$. Given that $\xi^{(E)}\propto \mu^2$ this shows how $\xi^{(A)}_{\mathrm{TMSV}} \propto \mu$, for $\mu\ll 1$. In such limit, 
 to obtain a desired $\epsilon$ and $\delta$ in Eq.~\eqref{eq:deltaeps}, 
it is sufficient to choose $M\propto (\ln\epsilon)^2/\delta$ and $\mu\propto
\sqrt{\delta/M} \propto |\delta/\ln\epsilon|$. 
Moreover, the total number of photons $\mu_T = \mu M \propto |\ln\epsilon|$ only depends on Alice error. For fixed $\mu_T$, and hence for fixed $\epsilon$, spreading the energy into more modes via a larger $M$ enables reducing $\delta$ to an arbitrarily small degree.

This theoretical analysis is corroborated by the numerical results shown in 
Fig.~\ref{fig:3}.\textbf{B}, which  gives a visualization of the main result of our paper, namely the achievability of quantum covert target ranging for sufficiently large $M$.
The classical counterpart is discussed in the End Matter and visualized in Fig.~\ref{fig:4}.{\bf B}. 
 
\section{Conclusions}\label{sec:Conclusions}
We analyzed the problem of covert target sensing in the optical domain by introducing a ``bottleneck'' on the Chernoff information. This bottleneck is determined by the requirement of hiding the probing party to an adversary, while also extracting information about the target. Solutions of the Chernoff information bottleneck define the covert 
information curve. We have shown that the shape of such a curve has all the information needed to assess the ability 
to perform covert sensing tasks in the asymptotic regime of many modes $M$. 
In the task of target ranging we showed how effective covert sensing is possible with quantum resources and impossible with classical ones.
Specifically we showed how using an entangled two mode squeezed vacuum state as a probe, one is able to perform covert target ranging with very low probability of error, whereas a coherent transmitter would recover  at best a very low amount of information while maintaining covertness.

Our work offers a novel approach in the investigation of sensing trade-offs and shows a very clear advantage in an important practical scenario in the task of target ranging. The measurements performed, namely photon counting, can be very well approximated by currently available single photons detectors. This means that our analysis has concrete relevance in near-term applications for LiDAR systems where low probability of detection is required. 

\begin{acknowledgments}
  The authors thank M.~Gu, J.~Thompson and S.~Pirandola for discussions. 
  The authors acknowledge financial support from:  
  European Union under the Italian National Recovery and Resilience Plan (PNRR) of NextGenerationEU, partnership on ‘Telecommunications of the Future’, PE00000001 - program ``RESTART'' (G.O.,~L.B.); 
  PNRR Ministero Universit\`a e Ricerca Project No. PE0000023-NQSTI, funded by European Union-Next-Generation EU (G.O.,~L.B.); 
  Prin 2022 - DD N. 104 del 2/2/2022, entitled ``understanding the LEarning process of QUantum Neural networks (LeQun)'', proposal code 2022WHZ5XH, CUP B53D23009530006 (L.B.); 
  the European Union’s Horizon Europe research and innovation program under EPIQUE Project GA No. 101135288 (L.B.);
 Defence Fund (EDF) under grant agreement 101103417 EDF-2021-DIS-RDIS-ADEQUADE, Funded by the European Union. Views and opinions expressed are however
those of the authors only and do not necessarily reflect those of the European Union or the European Commission. Neither the European Union nor the granting authority can be held responsible for them. 
\end{acknowledgments}

\bibliography{bib}

\newpage

\section*{End Matter}
\subsection{Chernoff Information for detection and ranging}
The Chernoff information is a central quantity in information theory \cite{Nielsen_2013}. In the classical setting 
it is a measure of distance between two probability distributions, $P_0(\mathbf{x})$ and $P_1(\mathbf{x})$, 
and it is defined as 
\begin{equation}
\xi(P_0,P_1)=\max_{0\leq \alpha \leq 1} C_\alpha(P_0,P_1), \label{eq:CIcla}
\end{equation}
where:
\begin{equation}
  C_\alpha(P_0,P_1)=-\log\left(\int d\mathbf{x}\, P^\alpha_0(\mathbf{x})P^{1-\alpha}_1(\mathbf{x})\right).
\end{equation}
In binary hypothesis testing, given $M$ samples from either distribution, the Chernoff information gives the optimal rate of exponential decay, in the asymptotic limit $M\to\infty$, of the probability of error.

The quantum counterpart of $\xi(P_0,P_1)$ is the Quantum Chernoff information \cite{Audenaert_2007} $\xi_{Q}$ between states $\rho_0$ and $\rho_1$:
\begin{equation}
\xi_{Q}(\rho_0,\rho_1)=\max_{0\leq \alpha \leq
1}-\log\left(\text{Tr}\left[\rho_0^\alpha\rho_1^{1-\alpha}\right]\right),
\end{equation}

which gives the best asymptotic decay rate of the probability of error in the discrimination of the multi-copy states $\rho_{i=0,1}^{\otimes M}$.
In this work we mostly use the classical Chernoff information between probability distributions defined as the result of photon-counting measurements, motivated by the choice of comparing  the quantum and classical performance in realistic remote sensing at optical frequencies, as discussed in the introduction. However, the formalism can be trivially extended to use $\xi_Q$ rather than $\xi$.

Alice's target sensing task, in the general case, can be expressed as an
$m-$hypoteses test. For a given probe state $\rho_0$, Alice will measure
the returning state $\rho_A$ and define hypothesis $\mathcal H^{(A)}_j$, with 
\begin{align}
  \mathcal{H}^{(A)}_j: \rho_A\stackrel?=\rho_{j}, &&
  j=1,...,m,
  \label{eq:alice hyp}
\end{align}
where $\rho_j=\mathcal{E}_{j,\kappa,\mu_B}(\rho_0)$ is one of the possible 
returning signals after the interaction with the target. The returning 
state depends on the hypothesis index $j$, on the reflectance 
$\kappa$ of the target and on the total number of background photons $\mu_B$. 

After probing the target with $M$ copies (modes), where Alice sends
$\rho_0^{\otimes M}$ and receives $\rho_{j}^{\otimes M}$, in the asymptotic regime, $M\gg 1$, the probability 
$p^{(A)}_{\mathrm{err}}$ that Alice guesses the wrong hypothesis
decays exponentially as  $p_{\mathrm{err}}^{(A)}\simeq e^{-M\xi^{(A)}}$, 
where the 
multi-hypothesis decay rate $\xi^{(A)}$ (both quantum and classical) is found in terms of that of 
the closest hypotheses in the set \cite{Nussbaum_2011, Li_2016}:
\begin{equation}
  \xi^{(A)} 
  =\min_{i,l}\left[ \xi(\mathcal{H}^{(A)}_i,\mathcal{H}^{(A)}_l)\right].
\end{equation}

On the other side, 
Eve faces a target detection task with no control on the states and the number of repetitions, which are chosen by Alice. 
This binary problem is defined by the hypotheses:
\begin{align}
&\mathcal{H}^{(E)}_0: \rho\stackrel?=\rho_B, & 
&\mathcal{H}^{(E)}_1: \rho\stackrel?= \mathcal{E}_{j,1-\kappa,\mu_B}(\rho_0),
\end{align}
where $\rho$ is the state being measured. In hypothesis $\mathcal H^{(E)}_0$ Eve receives 
a background state $\rho_B$, while in hypothesis $\mathcal H^{(E)}_1$ she receives 
the scattered input probe from Alice. Note that $\mathcal H^{(E)}_1$ is independent 
on the index $j$. 
Eve' best performance will be then characterized by the rate:
$\xi^{(E)} := \xi(\mathcal{H}^{(E)}_0,\mathcal{H}^{(E)}_1)$. 

\begin{figure}[t]
\includegraphics[width=0.5\textwidth]{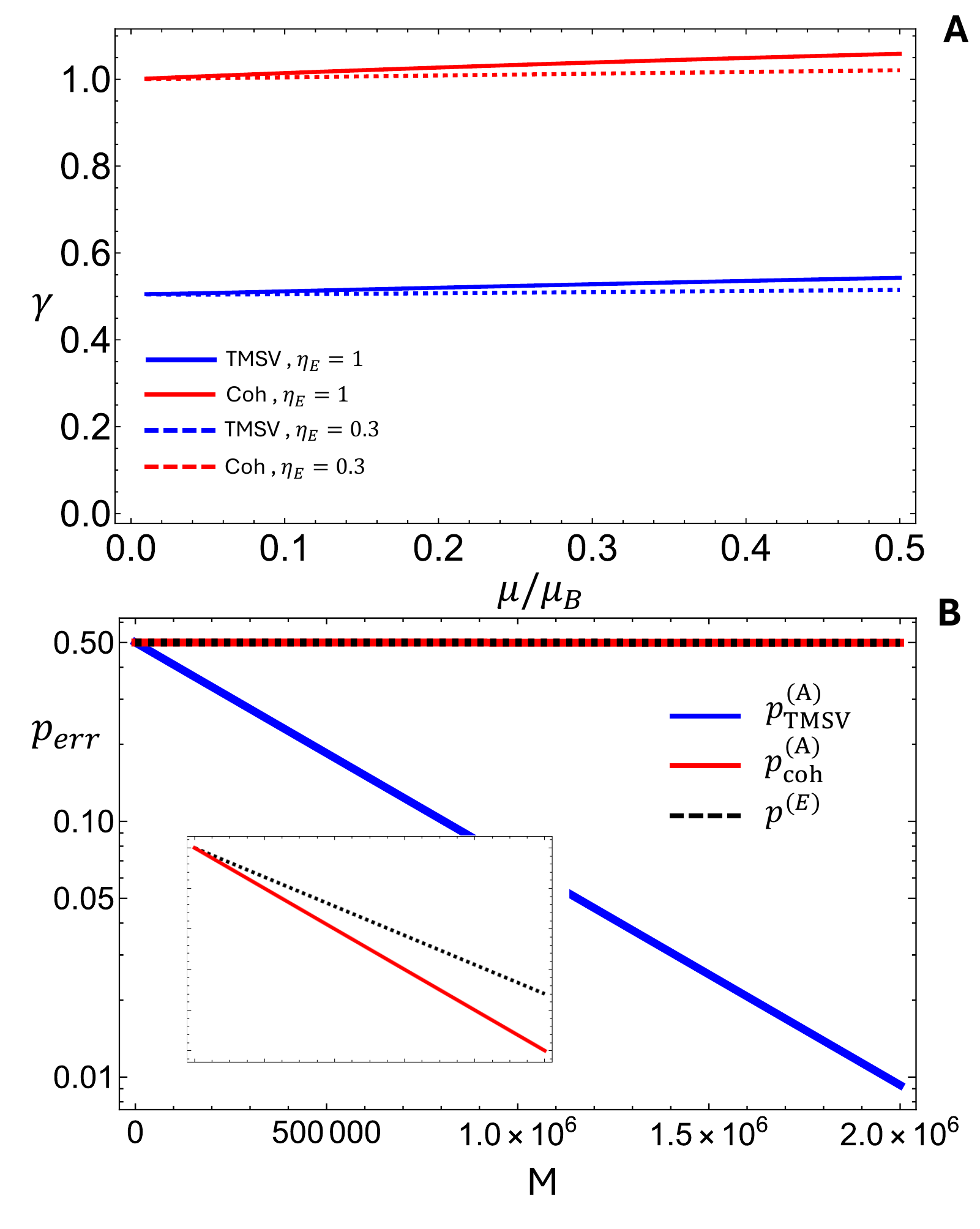}
	\caption{\textbf{A}. We plot the scaling parameter $\gamma$ as a function of $\mu/\mu_B$ (with fixed $\mu=0.01$). The parameters are the same of Fig.(\ref{fig:3}), with either $\eta_E=1$ (solid lines), or $\eta_E=0.3$ (dashed lines). \textbf{B}. Asymptotic probabilities of error (in log scale) with $\eta_E=0.3$ and the other parameters as Fig.(\ref{fig:3}). The insert it's a magnification to show the separate scaling of $p^{(E)}$ and  $p^{(A)}_{coh}$. }
  \label{fig:4}
\end{figure}
\subsection*{Photon counting measurements}
In our model we fix Alice measurements to phase insensitive ones, namely photon counting. This is because of our focus on the optical regime, in which the phase is hardly preserved due enviromental effects such as scattering and turbolence \cite{Ortolano_2024}. Fixing the measurement to
photon-counting means that the quantum hypothesis testing on the states is
translated in a classical testing on the photon number distributions.

Denoting the photon number distribution at the receiver when the target is present as  $P_{\kappa}$ and $P_B$ when it is not, then the target ranging Chernoff information is \cite{Ortolano_2024}:
\begin{equation}
\xi^{(A)}= 2 \mathcal{B}(P_{\kappa},P_B), \label{eq:CITR}
\end{equation}
where $\mathcal{B}(P_0,P_1):=C_{1/2}(P_0,P_1)$ is the Bhattacharyya information.

Since Eve operates in the same regime as Alice we fix the same conditions for both, i.e. same interaction model with target and background, and the same restriction to phase-insensitive measurements. Further justification for this latter restriction is given by the fact that Alice can modulate the probe phase, so that is effectively randomized to prevent Eve from extracting further information with phase measurements. Note that this modulation cannot be used by Alice to improve the scaling of the classical probe as phase information is still lost in the returning path. 
Eves best performance will be then characterized by the rate:
\begin{equation}\label{CI:E}
\xi^{(E)}= \max_{0\leq \alpha \leq 1} C_\alpha(P_{1-\kappa},P_B).
\end{equation}

 Alice's classical sensing, at a fixed number of total probe photons, can be improved, in principle, by sending less modes, and thus collecting less background noise. In the covert scenario, however, this strategy is not necessarily convenient, as this improvement would come at the cost of an increased probability of detection by Eve. Recalling that $p_{\rm err} \propto e^{-M \xi}$ when using $M$ copies of the probe state, we qualitatively analyze the alternative strategy with a large number of photons in a single mode via the quantity
\begin{equation}
\xi^{(A/E,\infty)}_{coh}= \lim_{M\to \infty} \xi^{(A/E)}_{coh}(M\mu=\mu',\mu_B)/M,
\end{equation}
which can be also understood as the asymptotic decay rate of the probability of error. 
From Eq.(\ref{eq:CIA}) we directly get $\xi^{(A,\infty)}_{coh}=\kappa\eta_A\mu'$, while $\xi^{(E,\infty)}_{coh}=(1-\kappa)\eta_E\mu'$ is computed from the exact expression of $\xi^{(E)}_{coh}$, since the approximated one from Eq.(\ref{eq:CIE}) does no longer hold for a large number of photons per mode.
In Fig.~\ref{fig:2} we study the performance of this alternative strategy for covert sensing, showing that it is close to the one obtained by spreading the photons in different modes. In view of this, in the main text we consider the latter as the main classical reference, though we will study both quantities.

\subsection{Classical Covert Sensing}
In this section we present additional results addressing the same set of parameters chosen for Fig.~(\ref{fig:2}.\textbf{A}-\textbf{B}), with focus on the scenario in which $\xi_{cla}^{A}>\xi^{E}$ showing even if this condition is met covert sensing cannot be performed efficiently.

The scaling parameter $\gamma$ (see main text) defines the condition of sublinear scaling for efficient sensing. For a classical probe we can expand $\gamma$ for low values of $\mu/\mu_B$ as:
\begin{equation}
\gamma= 1 + \frac{1}{4} \eta_B (1-2\kappa) \mu/\mu_B +\mathcal O(\mu^2/\mu_B^2). \label{eq:linscaling}
\end{equation}
Given that for $\kappa<1/2$ (faint target) the term $1-2\kappa>0$,  Eq.(\ref{eq:linscaling}) shows how in our regime of interest, $\mu\ll\mu_B$, $\gamma$ is always slightly larger then $1$, thus the classical strategy shows linear scaling. In Fig.~(\ref{fig:4}.\textbf{A}) $\gamma$ for the quantum (blue lines) and classical (red lines) probes as a function of $\mu/\mu_B$.  In the  regime of Fig.~(\ref{fig:2}.\textbf{A}), plotted in solid lines, $\xi_{cla}^{A}<\xi^{E}$. On the other hand for the parameters of Fig.~(\ref{fig:2}.\textbf{B}), plotted in dashed lines, $\xi_{cla}^{A}>\xi^{E}$ obtained by lowering Eves collection efficiency to $\eta_E<1$. However, the scaling parameter does not change significantly, showing similar behavior in both parameter settings. For the classical probe we have in both cases, $\gamma_{cla} \approx 1$, while for the quantum probe $\gamma_{qua} \approx 1/2$, preventing efficient covert sensing with the former and enabling it with the latter. 

This point is further shown in panel \textbf{B}, where we present the asymptotic probabilities of error when $\xi_{cla}^{A}>\xi^{E}$. The magnified insert shows that Alice, with a classical probe, is indeed recovering information at a greater rate then Eve. However the main figure shows how this rate is negligible for the purpose of covert sensing, that is indeed achieved by the quantum probe. 
\end{document}